\documentclass[twocolumn,showpacs,preprintnumbers,amsmath,amssymb]{revtex4}

\usepackage{graphicx}
\usepackage{bm}
 
\begin{document}

\title{Occupation times of  random walks in confined 
geometries: From random trap model to diffusion limited reactions.}

\author{S.Condamin}
\author{V.Tejedor}
\author{O. B\'enichou}
\affiliation{Laboratoire de Physique Th\'eorique de la Mati\`ere Condens\'ee 
(UMR 7600), case courrier 121, Universit\'e Paris 6, 4 Place Jussieu, 75255
Paris Cedex}

\date{\today}

\begin{abstract}
We consider a random walk in confined geometry, starting from a site and eventually 
reaching  a target site. We calculate analytically   the distribution of the 
occupation time on a third site, before reaching the target site. The obtained distribution is exact, and
completely explicit in the case or parallepipedic confining
domains. We discuss implications of these results in two different fields:  The mean first
passage time for the random trap  model  is computed  in dimensions greater than 1, and is shown to display
a non-trivial dependence with the  source and target positions ;  The probability of reaction with a given imperfect center before
being trapped by another one is also explicitly calculated, revealing
 a complex dependence  both in  geometrical and chemical
parameters.  
\end{abstract}


\maketitle


How many times, up to an observation time $t$,  a given site $i$ of a
lattice has been visited by a random walker? The study of  the
statistics of this  general  quantity,  known in the random walk
literature as the occupation time of this site has been  a subject of
interest for long,   both for mathematicians \cite{Aldous,Hughes} and
physicists \cite{Godreche,Majumdar,benichouJphysA2003,blanco,benichouepl2005,
condamin,blanco2,Barkai}. As a matter of fact,  the occupation time has
proven to be a key quantity  in various fields, ranging from
astrophysics \cite{ferraro2}, transport in porous media
\cite{Bouchaud} and 
diffusion limited reactions \cite{nouschimie}.  The point is that as soon as the sites of a
system have different physical or chemical properties, it becomes
crucial to know precisely how many times  each site is visited by the
random walker.

An especially important situation concerns the case when the
observation time $t$ up to which the occupation of site $i$ is
considered is itself random  and generated by the random walker. To
settle things and show how that occupation time ${\bf N_i}$ comes into play
in various physical situations,  we first give two  different
examples.

\begin{figure}[t]
\scalebox{0.8}{
\centering\includegraphics[width =.5\linewidth,clip]{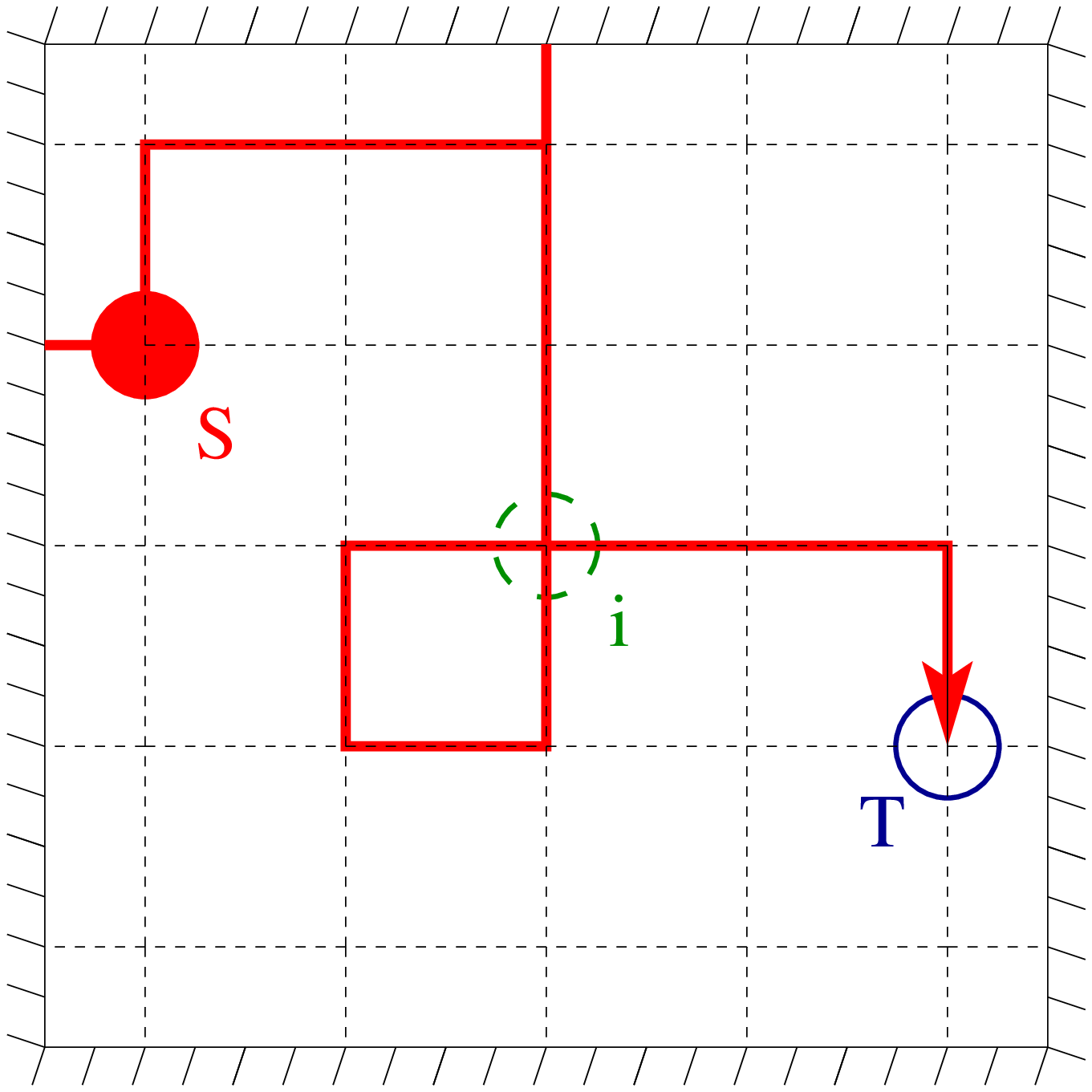}}
\caption{Color online. Schematic picture of the problem: the random walk begins at the 
site $S$, and the occupation time ${\bf N_i}$ is the number of times it 
visits the 
site $i$ before reaching the target $T$. In this picture, 
${\bf N_i} = 2$. 
}
\label{layout}
\end{figure}

The first one  concerns  the case of the so-called random trap model
(problem I), which is a very famous model of transport in quenched
disordered media \cite{Bouchaud}.  In this random trap model,  a
walker performs a symmetric lattice random walk, jumping towards
neighboring sites. In addition, the time the walkers spends at each
site is  a random variable $\tau_i$, drawn once and for all from a
probability distribution $\psi$, which is identical for all sites. 
A quantity which has proven
to be especially important in transport properties is the  first
passage time,  the time it takes to reach  a given target site. It is 
the key property in many physical applications \cite{Redner,LevitzPRL06}, ranging from
diffusion-limited reactions \cite{Blumen83,Yuste,Yuste2001,Yuste2003} to search processes 
(e.g. animals searching for food) \cite{nous}. The mean
first passage time (MFPT) for the random trap model has been studied
\cite{Murthy,Hernandez} but, to our knowledge, these determinations have been
strictly limited to the very specific 1D case, and higher dimensional
computations in confining geometries like in Fig.\ref{layout} are still
lacking (see nevertheless \cite{Kawasaki} for a $d$ dimensional related problem). The relation with the occupation time is the following:   The
MFPT at the target $r_T$ starting from site $r_S$  can be written down
as
$\displaystyle\langle {\bf T} \rangle=\sum_{i=1}^V \langle {\bf
N_i} \rangle \tau_i$,
 where  $V$ is the volume of the confining system, ${\bf
N_i}$ is the number of times the site $i$ has been visited before the
target is reached and $\langle ... \rangle$ stands for the average
with respect to the random walk. Concerning the distribution of the MFPT
with respect to the disorder, that is with respect
to the $\tau_i$'s, we are finally back to summing a
deterministic number $V$ of independent random variables    $\langle
{\bf N_i} \rangle \tau_i$ but non identically distributed (because of
the factor $ \langle {\bf N_i} \rangle$), which 
requires the determination of the {\it mean}
occupation times   $ \langle {\bf N_i} \rangle$ we introduced before.

The second situation has to deal with a very different problem (problem
II), which is involved for diffusion limited reactions in confined
media. We consider a free diffusing reactant $A$ that enters in a
cavity,  and which can react with a given fixed center $i$. We assume that each time the walker reaches the reactive site $i$,
it has a probability $p$ to react, which schematically  mimics an
imperfect reaction in confined geometry. Actually, numerous 
chemical reactions, ranging from trapping  in supermolecules 
\cite{KlafterJCP98} to activation processes of synaptic receptors 
\cite{Savtchenko, Holcman05b}  can be roughly rephrased by this generic 
scheme.  The question we address here, which does not seem to have been
considered before, is the  following: what is the probability for
 $A$ to react with the center $i$ before exiting the
cavity? More generally, for a random walker starting from a site
$S$, what is the probability $Q$ to react with $i$ before
reaching a target site  $T$, possibly different from
$S$. Partitioning over the number of times the reactive site $i$
has been visited, we have:
\begin{equation} \label{QQ}
Q=1-\sum_{k=0}^\infty P({\bf N_i}=k)(1-p)^k.
\end{equation} Once again, the random variable ${\bf N_i}$ is
involved, but that time the determination of the {\it entire
distribution}  $P({\bf N_i}=k)$ is needed.

In this Letter, we propose for the first time a method of
computation of the statistics of ${\bf N_i}$ in confining geometry. In
particular, we  obtain explicitly the {\it exact distribution} in the
case of parallepipedic confining domains. Applications to the above
mentioned examples are discussed.

 We start with the computation of the mean $ \langle {\bf N_i}
\rangle$, assuming for the time being that the starting and target sites are different ($S\neq T$). We note $w_{ij}$ the transition probabilities from site $j$
to site $i$. We have $\sum_i w_{ij} = 1$, and we take $w_{ij} =
w_{ji}$.  These general transition probabilities can take into account
reflecting  boundary conditions.
 We consider an outgoing flux $J$ of
particles in $S$. Since the domain is finite, all the particles are
eventually absorbed in $T$, and, in the stationary regime, there is
an incoming flux $J$ of particles in $T$. The mean particle density
$\rho_i$ thus satisfies the following equation:
\begin{equation} \rho_i = \sum_j w_{ij}\rho_j + J\delta_{iS} -
J\delta_{iT},
\label{eqni} 
\end{equation} 
with the boundary condition $\rho_T = 0$ (it
is the absorbing site). To find the mean occupation time, we can simply
notice that the mean particle density $\rho_i$ is equal to  
$\langle {\bf N_i} \rangle J$. 
To solve this problem, we use the pseudo-Green
function $H$ \cite{Barton,PRE2006}, which satisfies:
\begin{equation} H({\bf r}_i|{\bf r}_j) = \sum_k w_{ik} H({\bf
r}_k|{\bf r}_j) + \delta_{ij} -  \frac1V,
\label{defH}
\end{equation} where $V$ is the total number of sites of the
lattice. It is also symmetrical  in its arguments,
and the sum $\displaystyle \sum_i H({\bf r}_i|{\bf r}_j)$   
is a constant independent of $j$. Using the concise notation $H_{ij} = H({\bf
r}_i|{\bf r}_j)$, 
it can be seen by direct substitution that $\rho_i$ given by
\begin{equation} \langle {\bf N_i} \rangle = \frac{\rho_i}{J}=H_{iS} - H_{iT} + H_{TT}
- H_{ST},
\label{resni}
\end{equation} 
satisfies Eq.(\ref{eqni}) as well as the boundary condition $\rho_T=0$. 
 Note that 
these results also give the mean occupation time of a subdomain, which is 
simply the sum of the mean occupation time of all the sites in the subdomain. 
In particular, we can check that the mean occupation time for the whole 
domain, 
$\displaystyle\sum_{i=1}^V \langle {\bf N_i} \rangle = V(H_{TT} - H_{ST})$, 
gives back  the MFPT
from $S$ to $T$ \cite{CondaminPRL,PRE2006}.

Before we go further, it is necessary to give a few elements on the
evaluation of $H$ for isotropic random walks. The following exact expression \cite{CondaminJCP,PRE2006}
is known in two dimensions for rectangles: 
\begin{eqnarray}\label{Hexact}
H({\bf r}|{\bf r}') = \frac4N \sum_{m=1}^{X-1} \sum_{n=1}^{Y-1}
\frac{\cos\frac{m\pi x'}{X}\cos\frac{n\pi y'}{Y}
\cos\frac{m\pi x}{X}\cos\frac{n\pi y}{Y}}{
1-\frac12\left(\cos\frac{m\pi}{X}+\cos\frac{n\pi}{Y}\right)}\nonumber \\
+ \frac4N \sum_{m=1}^{X-1}\frac{\cos\frac{m\pi x'}{X}\cos\frac{m\pi x}{X}}{
1 - \cos\frac{m\pi}{X}}  + 
\frac4N \sum_{n=1}^{Y-1}\frac{\cos\frac{n\pi y'}{Y}\cos\frac{n\pi y}{Y}}{
1 - \cos\frac{n\pi}{Y}},
\end{eqnarray}
where $X$ and $Y$ are the dimensions of the rectangle, and the coordinates 
$x$ and $y$ are half-integers going from $1/2$ to $X-1/2$ or $Y-1/2$. There 
is also a similar expression for parallepipedic domains in three dimensions. 
In more general domains, the most basic approximation (which usually gives a
good order of magnitude) is to approximate $H$ by the infinite-space
lattice Green function $G_0$ \cite{Barton},  $G_0$ being evaluated as
$G_0({\bf r}|{\bf r}') = 3/(2\pi|{\bf r}-{\bf r}'|)$ for  ${\bf r}
\neq {\bf r}'$, and $G_0({\bf r}|{\bf r}) = 1.516...$ in three
dimensions, and $G_0({\bf r}|{\bf r}') = -(2/\pi)\ln|{\bf r}-{\bf
r}'|$ for  ${\bf r} \neq {\bf r}'$, and $G_0({\bf r}|{\bf r}) =
1.029...$ in two  dimensions.  
More accurate approximations can be found \cite{PRE2006}, 
but the  above approximations are good enough to capture the qualitative 
behavior of the pseudo-Green
function, and of the distribution of the occupation time.

It is indeed possible to obtain not only the mean, but also the entire
distribution of the occupation time. The idea to tackle this {\it a priori}
difficult problem is to   use recent
results concerning the so-called splitting
probabilities \cite{Redner,CondaminPRL,PRE2006}. In presence of two targets $T_1$ and $T_2$, the
splitting probability $P_1$ to  reach $T_1$ before $T_2$ is \cite{CondaminPRL,PRE2006}:
\begin{equation} P_1 =
\frac{H_{1S}+H_{22}-H_{2S}-H_{12}}{H_{11}+H_{22}-2H_{12}}
\end{equation}
Denoting here $P_{ij}(i|S)$ the  splitting probability to reach
$i$ before $j$,  starting from $S$,   we have $P({\bf N_i} =0)= P_{iT}(T|S)$, and  for $k \geq 1$: 
\begin{equation} P({\bf N_i}=k) = P_{iT}(i|S)\left[\sum_j w_{ji}
P_{iT}(i|j)\right]^{k-1} \left[\sum_j w_{ji} P_{iT}(T|j)\right]
\end{equation} The three terms of this last equation correspond respectively to the
probability to reach $i$  before $T$, starting from $S$, the
probability to return to $i$ before  reaching $T$, starting from $i$,
to the power $k-1$, and the probability to reach $T$ before returning
to $i$. It can thus be written
\begin{equation} \label{1} P({\bf N_i}=k) = AB(1-B)^{k-1}\;\;{\rm
for}\;\;k\ge 1,
\end{equation} with
\begin{equation} \label{2} A \equiv P_{iT}(i|S) =
\frac{H_{iS}+H_{TT}-H_{ST}-H_{iT}}{H_{ii}+H_{TT}-2H_{iT}},
\end{equation} and
\begin{eqnarray}
B & \equiv & \sum_j w_{ji} P_{iT}(T|j) = 1 - \sum w_{ji} P_{iT}(i|j)\\
& = & \frac{\sum_j w_{ji} H_{Tj}-H_{iT}-\sum_j w_{ji} H_{ji}+H_{ii}}{H_{ii}+H_{TT}-2H_{iT}}\\
& = & \frac{1}{H_{ii}+H_{TT}-2H_{iT}},\label{3}
\end{eqnarray}
using Eq.(\ref{defH}), and $\sum_i w_{ij} = 1$. It can also be noted that 
$P(\mathbf{N}_i = 0) = 1 - A$.  The distribution of the occupation numbers
 given by Eqs.(\ref{1})-(\ref{3}) is the main result of
this Letter, and several comments are in order.
(i) Expressions of $H$  given in Eq.(\ref{Hexact}) makes this result  exact and completely explicit 
for parallepipedic domains.
(ii) Computing $\langle {\bf N_i} \rangle$ with this
distribution gives back the expected result (\ref{resni}).
(iii) It can be noted here that $B$, which characterizes the
decay of the probability distribution of ${\bf N_i}$, is independent
of the source. In addition, qualitatively, the basic evaluations of
$H$ following Eq.(\ref{Hexact}) (namely $H = G_0$) give for $B$ the following 
order of magnitude, if 
$i$ and $T$ are at a distance $R$: 
\begin{equation}
B \simeq \left\{
\begin{array}{ll}
\left[2G_0(0) - 3/(\pi R)\right]^{-1} & \; \mathrm{in}\; \mathrm{3D},\\
\left[2G_0(0) + (4/\pi)\ln R\right]^{-1} & \; \mathrm{in}\; \mathrm{2D},
\end{array}
\right. 
\label{23Dapprox}
\end{equation}
where $G_0(0) = G_0({\bf r}|{\bf r})$ is a dimension-dependant constant, given
in the discussion on the evaluation of $H$. 
This shows that $B$  decreases with the distance
between $i$ and $T$: a larger distance corresponds to a  slower decay. But, 
while it tends towards $0$ in two dimensions (which  corresponds to a wide
distribution of ${\bf N_i}$, and a large variance), it tends  to a
finite value in three dimensions. It can thus be said that  the sites
much further from the target than the source have,  in three
dimensions, a significant probability to be visited, but a  low
probability to be visited many times, whereas, in two dimensions,
they have a low probability to be visited at all, but a comparatively
high  probability to be visited many times. This is connected with the
transient  or recurrent character of the free random walk in two or
three dimensions.
(iv) The results obtained here for different starting and
target sites may easily be adapted to identical starting and target
sites ($S=T$):
\begin{equation} \label{special}
P({\bf
N_i}=0)=1-B\;;\; P({\bf N_i}=k) = B^{2}(1-B)^{k-1} \;\;{\rm
for}\;\;k\ge 1.
\end{equation} 
Note that this gives in particular a mean occupation
time of $1$ for all sites, a result which could  be derived
from an extension of Kac's formula \cite{PRE2006,Aldous}. However, here, we obtain not
only the mean occupation number but the entire distribution of this
occupation number, which appears to  vary from site  to site: the
further the site is from the target, the slower the probability
distribution decays.

We now discuss the applications of these general results to the
examples mentioned  in the introduction.  As for the random trap
model (problem I), we focus here on the  especially interesting case
of a  one-sided Levy stable distribution \cite{Hughes} $\psi(t) =
f_\alpha(t,\tau_0\alpha\cos(\pi\alpha/2),1,0))$ ($0<\alpha<1$), which
corresponds to an algebraic decay:
\begin{equation} 
\psi(t) \sim \frac{\alpha
\tau_0^\alpha}{\Gamma(1-\alpha)t^{1+\alpha}}
\label{defpsi}
\end{equation} 
and whose Laplace transform is $\hat{\psi}(u)= \exp(-\tau_0^\alpha u^\alpha)$
 ($\tau_0$ can be seen as the typical waiting time).
The Laplace transform $\hat{\pi}(u)$ of the distribution of the MFPT with respect to the
disorder reads
\begin{equation} 
\hat{\pi}(u) = \prod_{i=1}^V \hat{\psi}( \langle
{\bf N_i} \rangle u)=\exp(-(T_{typ} u)^\alpha)
\end{equation} 
The probability density of the MFPT is then as
could have been expected a one-sided Levy stable law, but    with a
non trivial typical time:
\begin{equation} T_{typ} = \tau_0\left( \sum_{i=1}^V (H_{iS} - H_{iT} +
H_{TT} - H_{ST})^\alpha \right)^{1/\alpha}
\label{ttyp}
\end{equation} 
For large size domain $V$, this result can be applied to  any
wide-tailed distribution of the waiting times satisfying
Eq.(\ref{defpsi}) \cite{Bouchaud}. 
It can be shown that $T_{typ}$ is bounded by 
$\tau_0 V^{1/\alpha}(H_{TT} - H_{ST})$, and tends towards this upper bound as 
$V$ grows, which provides  a simple estimation of $T_{typ}$ and indicates that  for large enough domains, the scaling of $T_{typ}$ 
with the source and  target positions is the same as for the 
discrete-time random walk (pure systems) \cite{CondaminPRL,PRE2006}. 
We thus showed that the random trap problem in confined geometries,
with a  wide-tailed waiting time distribution, has a Levy distribution of mean
first-passage times,  with a non-trivial typical time. The scaling
with the size $V$ is $V^{1/\alpha}$. The scaling with the source
and target positions is modified by the disorder in small confining
domains, while it is the same  as for pure systems
 in large enough domains.

Concerning the application to diffusion-limited reactions (problem II), the probability
$Q$ to have reacted with $i$ before reaching $T$ writes, using 
Eqs.(\ref{QQ}),(\ref{1}): 
\begin{equation}\label{eqQQ}
Q=\frac{Ap}{1-(1-p)(1-B)}
\end{equation}

\begin{figure}[t]
\scalebox{0.8}{
\centering\includegraphics[width =.7\linewidth,clip]{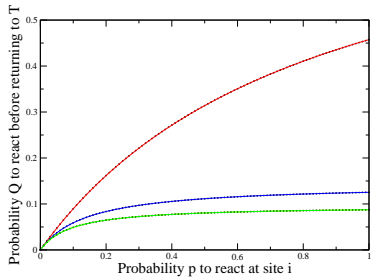}}
\caption{Color online. Simulations (symbols) versus analytical prediction (lines) Eq.(\ref{eqQQ}) of the probability $Q$ to react before returning to the target site as a 
function of the probability to react at site $i$. The confining domain is 
a square of side 51, and the target is at the middle of an edge, of 
coordinates (0,25), the site (0,0) being a corner site. The three curves 
corresponds to different positions of site $i$: (1,25) (red upper curve), 
(25,25) (blue mid curve), and (50,25) (green lower curve). }
\label{Q}
\end{figure}

The expression (\ref{eqQQ}) displays a subtle interplay between the
geometrical factors, involved through the terms $A$ and $B$, and the
reactivity $p$. Focusing now on the specific case of identical
starting and target points (meaning $A = B$, cf. Eq.(\ref{special})),
we exhibit two interesting limiting regimes.  In the ``reactivity
limited regime'', defined by   $p \ll B$,   we have $Q \sim p$. In
particular, in that regime $Q$  does  not depend on the reactive site
$i$. In other words, for a fixed reactivity $p$, all sites $i$ such
that  $p \ll B$ have the same probability of reaction $Q$, and the
detailed position of $i$ does not come into play.  On the
contrary, the ``geometrically limited regime''  $p \gg B$ leads to $Q
\sim B$, which does no longer depend on $p$, but only on the geometry.
Given the order of magnitude of $B$ (cf. Eq. (\ref{23Dapprox})), 
this can essentially happen in two dimensions, 
when $R \gg \exp(-\pi/(2p))$. This can be explained by the recurrent 
character of the two-dimensional random walk: when the reacting site 
$i$ is far enough from the target, if a random walker reaches it, it 
is likely to visit it many times before returning to $T$, and is 
thus  almost sure to react whenever  $i$ is reached. The reaction
probability $Q$ then becomes the probability to reach the site $i$. 
Consequently, the
position of the reacting site has a low influence on reactivity in three 
dimensions, or  when the reacting site is within a disk of 
radius $R = \exp(-\pi/2p)$ around the target in two dimensions. If the reacting site is further,
the geometrical effects become preeminent. 
We 
show in  Fig. \ref{Q} a graph of $Q$, as a function of $p$, for  different
positions of $i$ (near the target, in the middle of the  domain, and
at the opposite),  the source and target point being identical. The
limiting regimes can  be well identified.

To conclude, we have computed the  distribution  of the occupation
time of a given site $i$, for a random walk in confined geometry,
eventually trapped at a target.  This distribution is exact, and
completely explicit in the case or parallepipedic confining
domains. While the mean occupation time, unsurprisingly,  is higher
when $i$ is near the source and lower near the target (and uniform if
the  source and target are identical), the distribution of the
occupation time is  essentially exponential, with a slower decay when
the point is far away from the target. We have also presented
important applications of these results in two different fields. The
first one is transport  in quenched disorder media: The mean first
passage time for the random trap  model  has been computed  for the
first time in dimensions greater than 1, and has been shown to display
a non-trivial dependence with the  source and target positions. The
second application is to diffusion limited reactions in confined
geometry: The probability of reaction with a given imperfect center before
being trapped by another one has been explicitly calculated, and has proven to
present a complex dependence  both in the geometrical and chemical
parameters.   We believe that the results obtained in this Letter
could be relevant to   systems  involving diffusion in  confining
domains, displaying inhomogeneous physical or chemical properties.  .

\end{document}